\documentclass[10pt,twocolumn,twoside, amsmath, amssymb, aps]{revtex4-2}
\usepackage{soul}
\usepackage{graphicx,bm,hyperref}% Include figure files
\usepackage[mathlines]{lineno}% Enable numbering of text and display math
\usepackage{xcolor}
\begin{document}

\title{Protein-DNA Co-condensation is Prewetting to a Collapsed Polymer}
\author{Mason N. Rouches}
\email{mason.rouches@yale.edu}
\affiliation{Department of Molecular Biophysics \& Biochemistry, Yale University}
\affiliation{Quantitative Biology Institute, Yale University}
\affiliation{Department of Physics, University of Chicago}
\author{Benjamin B. Machta}
\email{benjamin.machta@yale.edu}
\affiliation{Department of Physics, Yale University}
\affiliation{Quantitative Biology Institute, Yale University}

\begin{abstract}
The three-dimensional organization of chromatin is thought to play an important role in controlling gene expression. Specificity in expression is achieved through the interaction of transcription factors and other nuclear proteins with particular sequences of DNA. At unphysiological concentrations many of these nuclear proteins can phase-separate in the absence of DNA. In-vivo the thermodynamic forces driving these phases lead the chromosome to co-condense with nuclear proteins. However it is unclear how DNA, itself a long polymer subject to configurational transitions, interacts with three-dimensional protein phases. Here we show that a long compressible polymer can be coupled to interacting protein mixtures, leading to a generalized prewetting transition where polymer collapse is coincident with a locally stabilized liquid droplet. We use lattice Monte-Carlo simulations and a mean-field theory to show that these phases can be stable even in regimes where both polymer collapse and coexisting liquid phases are unstable in isolation, and that these new transitions can be either abrupt or continuous.  For polymers with internal linear structure we further show that changes in the concentration of bulk components can lead to changes in three-dimensional polymer structure. In the nucleus there are many distinct proteins that interact with many different regions of chromatin, potentially giving rise to many different Prewet phases. The simple systems we consider here highlight chromatin's role as a lower-dimensional surface whose interactions with proteins are required for these novel phases. \\~\\
\fbox{\parbox{0.77\textwidth}{\noindent\textbf{Significance Statement} --- Proteins and DNA in the nucleus display rich spatial organization, but the forces which drive it are not well understood. Here we show that proteins prone to forming condensed liquid droplets can drive configurational phase transitions of long polymers, like DNA, even when too dilute to phase separate on their own. Indeed, many transcription factors (TFs) will condense into liquid phases in the absence of DNA when enriched to much higher concentrations than in the nucleus. With DNA, and at much lower TF concentrations, we expect these proteins to undergo generalized prewetting transitions, leading to abrupt changes in the three dimensional organization of chromatin. We argue that these phase transitions play an important role in organizing and regulating chromatin.}}
\end{abstract}

\maketitle

In the nucleus of eukaryotes chromosomes display three dimensional structure at many distinct scales~\cite{Lieberman2009Comprehensive}. In some cases the co-localization of enhancers, promoters, and otherwise distal genetic elements has been shown to be a key step in initiating transcription~\cite{du2024direct,NolisTranscription2009,zuin2022nonlinear}. While some of this structure is driven by enzymatic activity from loop extrusion factors (LEFS) and polymerases ~\cite{GoychukPolymer2023,fudenberg2016formation}, some is likely to be thermodynamic in origin, driven by energetic interactions between long polymers and proteins, small RNAs and other macromolecules~\cite{SabariCoactivator2018,strom2017phase,shrinivas_enhancer_2019}. Many nuclear components have a tendency to phase separate into coexisting liquid phases when isolated at high concentrations, even in the absence of DNA~\cite{larson_liquid_2017,SabariCoactivator2018,morin_sequence-dependent_2022}. These protein phases likely co-condense with chromatin to organize the genome ~\cite{gibson_organization_2019,sharma_liquid_2021,du2024direct,basu_unblending_2020,strom_interplay_2024}.

%(3) Polymer Physics: explain the polymer collapse transition, how it is modulated by a goodness/badness of solvent.  motivate that it too seems plausibly related to chromosome organization
But there are other thermodynamic drivers of spatial organization available to long polymers like chromosomes. In particular, Collapse transitions separate a regime where typical configurations are extended from one where polymers are condensed and space filling~\cite{de1975collapse}. These transitions can be accessed by solvent quality; in good solvents polymers typically have a weak contact repulsion and exist in an Extended configurational phase. In poor solvents, polymers prefer to maximize self-contacts, leading to Collapsed phases. In the nucleus more complex interactions could in principle drive more interesting configurational phases.

%(4) motivate the idea of a coupled transition.  Talk about our work in membranes, work in synaptic vesicles, etc. etc.
We recently argued that a range of structures at the plasma membrane are \textit{Prewet}: surface phases formed by a combination of membrane mediated forces and interactions between cytoplasmic proteins~\cite{rouches_prewet_21}. In classical prewetting, a thin film resembling a three dimensional phase can adhere to a surface outside of a thermodynamic regime where that bulk phase is stable~\cite{cahn_critical_1977,nakanishi_multicriticality_1982}.  But in the biological case, the solid surface is replaced by the fluid plasma membrane, itself found near a 2D liquid-liquid demixing critical point~\cite{veatch_critical_08}. In this system the prewetting transition of the bulk merges with the de-mixing transition at the surface, allowing for phase transitions that would not occur in the membrane or bulk alone.

Here we explore a model in which bulk proteins prone to phase separation play an analogous role in modulating the properties of a polymer collapse transition. In this \textit{generalized prewetting} transition there is a phase in which bulk components are at near condensed density, and in which the polymer is Collapsed, at parameters in which the polymer would remain Extended in the absence of bulk, and the bulk would remain well-mixed in the absence of a polymer. We investigate the phase diagram of this generalized prewetting transition with both a lattice model and a mean field theory.

\begin{figure*}[ht!]
    \centering
    \includegraphics[width=13.4cm]{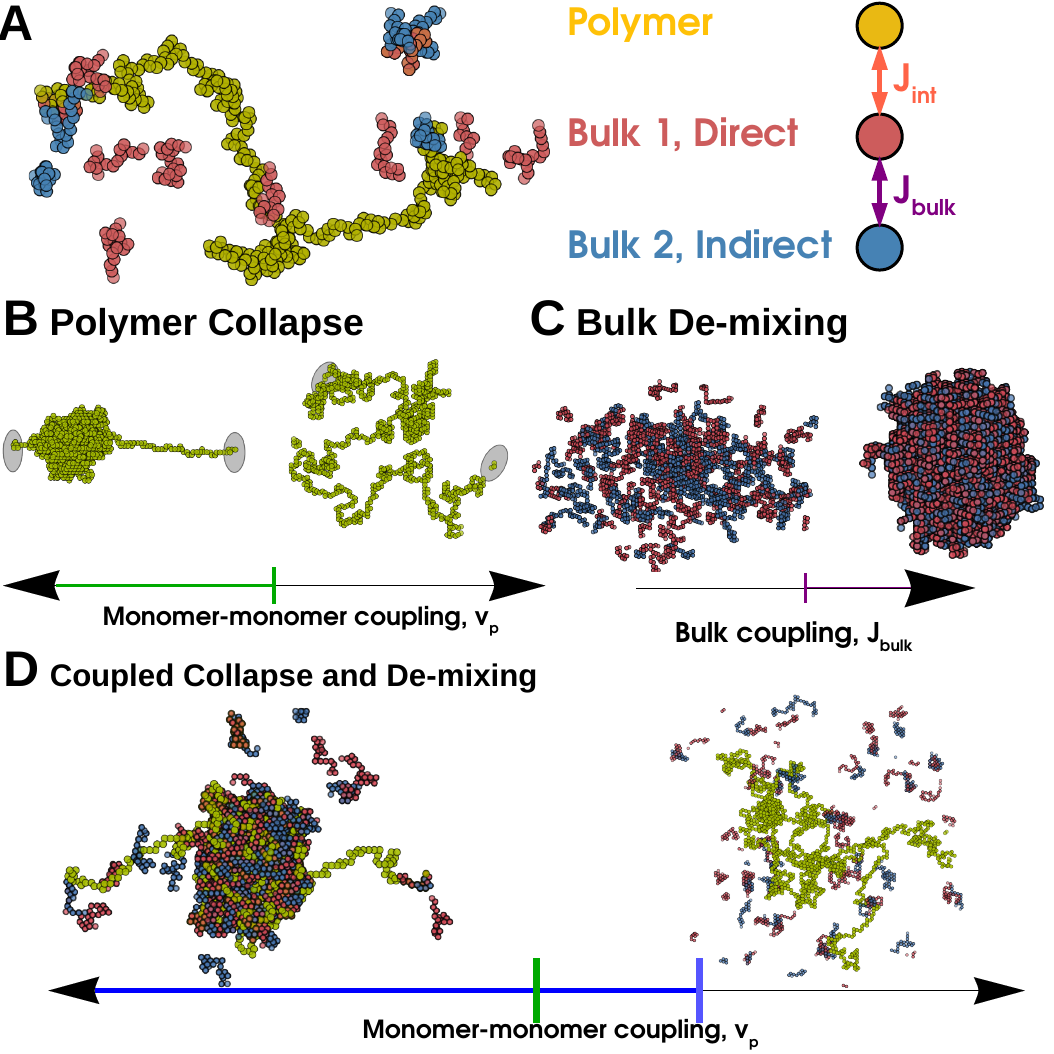}
    \caption{\textbf{Polymer collapse and phase-separation are coupled in a Generalized Prewetting phase transition} --- (A) Model coupling polymer collapse and liquid-liquid phase-separation. The long polymer, yellow, interacts with the bulk, red and blue molecules. The polymer is self-avoiding but its monomers have nearest-neighbor interactions that can be attractive or repulsive; the red and blue bulk molecules interact with each other with energy $J_{\text{bulk}}$. The red molecule and the long polymer interact with energy $J_{\text{int}}$ and the blue molecule does not interact with the long polymer. (B) A long polymer undergoes a Collapse transition where the polymer's configuration changes rapidly but continuously when monomer-monomer coupling  changes sign, green line. (C) The 3D bulk de-mixes through a first-order phase transition driven by interaction (or composition). (D) When polymer and bulk systems interact, the collapse transition and bulk demixing coincide in a Generalized Prewetting transition, illustrated below as the blue line, which occurs at monomer-monomer couplings more repulsive than those required for collapse in the absence of bulk molecules. The corresponding uncoupled system would have a single Dilute bulk phase and an Extended polymer.}
    \label{fig:fig1}
\end{figure*}

\section*{Monte-Carlo Simulations} \noindent\textbf{Model:} We model a single self-avoiding polymer on a $3$-D cubic lattice of linear dimension $L=64$. A spin variable at lattice site $i$, $s^p_{i}=1$ for sites which the polymers passes through and is $0$ otherwise. The energy of configurations of this system is given by the Hamiltonian $H_{\text{poly}}$:
\begin{equation}
\frac{H_{\text{poly}}}{k_{\text{B}}T} = -(\mu_{\text{p}}+v_{\text{p}})\sum_{i} s^{\text{p}}_{i} + v_{\text{p}}\sum_{<i,j> }s^{\text{p}}_{i} s^{\text{p}}_{j}
\end{equation}
Here both sums are over lattice sites, and $<i,j>$ represents nearest neighbors. $\mu_p$ is a monomer chemical potential which multiplies $N_{\text{p}}=\sum_i s_i^{\text{p}}$, the total number of monomers in the polymer. $v_{\text{p}}$ is a monomer-monomer interaction energy, and the subtraction in the first term avoids counting contacts between sequential monomers.  All simulations take place on periodic lattices, and are constrained to polymer configurations with initial position $(L/2,L/2,0)$ and final position $(L/2,L/2,L)$.  We also enforce that the polymer have length shorter than $N_{\text{max}}$, typically $1500$ monomer units, so that the polymer obeys $L \leq N_{\text{p}}<N_{\text{max}}$. In most of our simulations $N_{\text{p}}\approx N_{\text{max}}$, so that we sample an effectively fixed length distribution, but allow small fluctuations for numerical convenience. 

%(2) Describe lattice Hamiltonian for bulk, again explaining that it corresponds to TFs mediator, HP1, whatever, and then also explain the interaction between the bulk and the long polymer.
We couple this single long polymer (yellow in Fig.\ref{fig:fig1}A) to a solution of two types of shorter polymers which we term \textit{bulk molecules} (red and blue in Fig.\ref{fig:fig1}A). Bulk molecules are held at fixed chemical potential $\mu_{\text{b}}$ and fixed length $N_{\text{b}}$. 
Following our previous work~\cite{rouches_prewet_21}, we model the bulk as a mixture of two types of molecules prone to condensing together.  Molecules of the same type are self-avoiding while molecules of different types may occupy the same site, and have an on-site attractive interaction with energy $J_{\text{bulk}}$. Their occupancies at site $i$ are defined analogously to the long polymer by two additional spin variables $s^{1}_{i}$ and $s^{2}_{i}$. All molecules interact with a weak nearest-neighbor energy $J_{\text{nn}}$. The bulk Hamiltonian thus reads:
% Bulk length is Nb, chemical potential one over nb
\begin{align}
\frac{H_{\text{bulk}}}{k_{\text{B}}T} = &-\frac{\mu_{\text{b}}}{N_{\text{b}}}\sum_{i} \left(s^{1}_{i} + s^{2}_{i}\right)-J_{\text{bulk}}\sum_{i} s^{1}_{i}s^{2}_{i} \nonumber\\ &- J_{\text{nn}}\sum_{<i,j>}\left(s^{1}_{i}s^{1}_{j} + s^{1}_{i}s^{2}_{j} +s^{2}_{i}s^{2}_{j}\right) 
\end{align}

%% Fig 3: coupled d coupled phases

The long polymer and one of the bulk molecules, red in the Figures, are coupled by interactions of strength $J_{\text{int}}$, giving an interaction energy:
\begin{equation}
\frac{H_{\text{int}}}{k_{\text{B}}T} = -J_{\text{int}}\sum_{i} s^{1}_{i} s^{\text{p}}_{i}
\end{equation}
We illustrate this model in Fig.\ref{fig:fig1}A. 

%(3) Briefly explain the MCMC procedure
We use a Monte-Carlo procedure to sample equilibrium configurations of this system (see Methods). We sample the long polymer through three elementary moves that add, delete, or `kink' a given bond. We exchange bulk molecules with a particle reservoir and update their configuration with moves that conserve length.\\  %We accept moves with Metropolis probability chosen to obey detailed balance given the above Hamiltonian.\\
\begin{figure*}[htb!]
    \centering
    \includegraphics[width=7in]{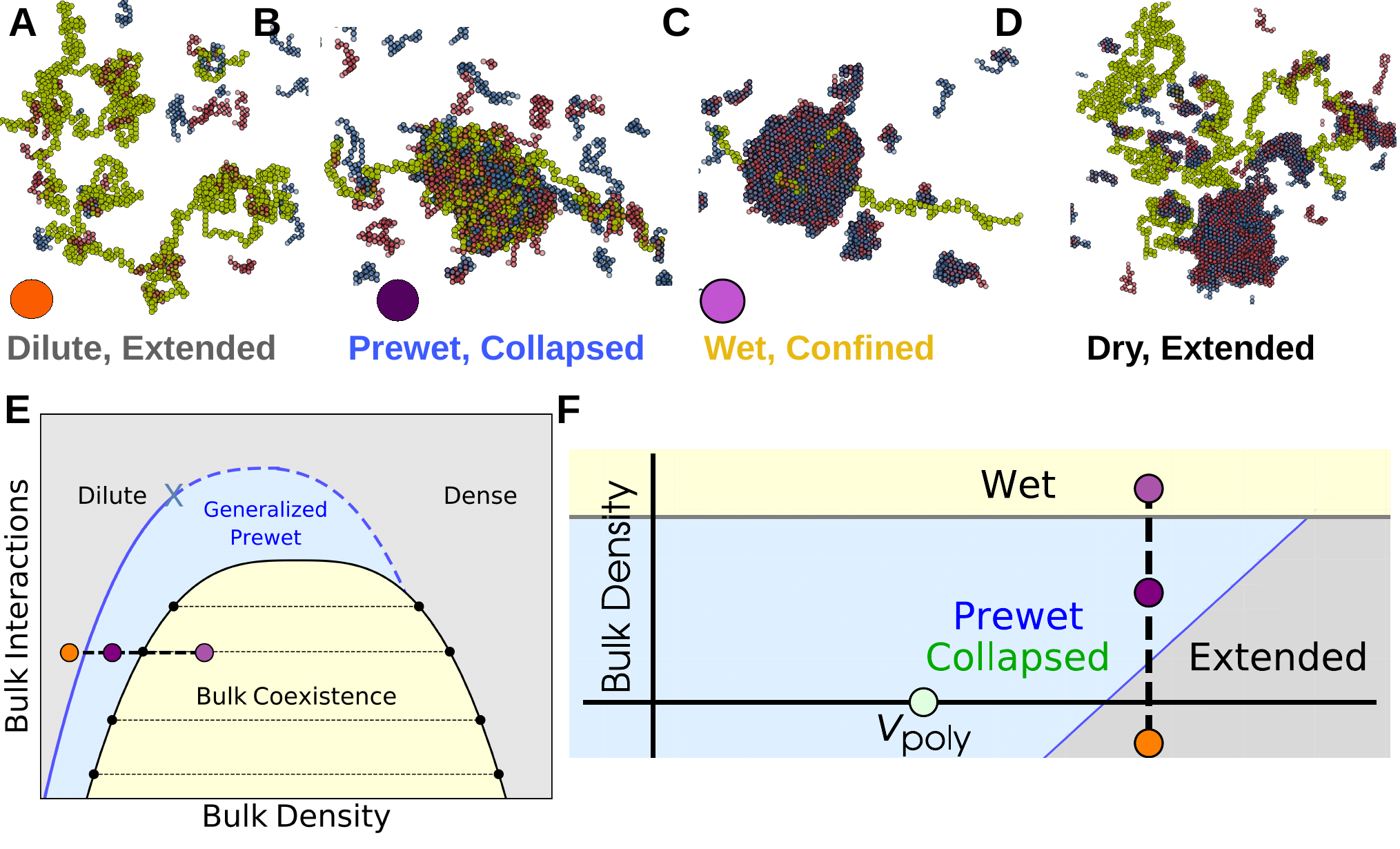}    
    \caption{\textbf{The bulk system shifts the Polymer Collapse phase boundary} --- (A-D) Configurations of coupled phases: Outside of the bulk coexistence region there are two phases that correspond to phases of an isolated polymer: (A) Extended and (B) Collapsed. The Collapsed phase coincides with condensation, or Prewetting, of the bulk fluid. The Extended phase has corresponding Dilute bulk phase. Within the bulk coexistence region the polymer may prefer to enter the droplet, confining itself (Wet, C) or avoiding it (Dry, D). In C and D we hold bulk particle number fixed to observe bulk coexistence. (E) Schematic bulk phase diagram: Bulk coexistence of a dense and dilute phase occurs in the yellow shaded region.  Generalized Prewet phases that coincide with polymer collapse occur outside of the bulk coexistence region (light blue region). 
 These can be accessed by increasing bulk density through a first order transition (solid blue line) or through a continuous transition (dashed blue line). These lines meet at a tricritical point (blue x). %and \st{terminate in a critical point above the bulk critical temperature}.
 Colored points correspond to above configurations. (F) Schematic phase diagram of monomer-monomer interactions $v_{\text{p}}$ and bulk chemical potential $\mu_{\text{b}}$. The bulk modifies the phase boundary of the polymer, allowing it to collapse in regions where it otherwise would not; this Collapse transition coincides with condensation of the bulk on the surface of the polymer.}
    \label{fig:fig2}
\end{figure*}
\noindent\textbf{Thermodynamics of an Isolated Polymer} ---
%(1) What is a polymer phase? Explain what a phase is and how polymers have phases that are not really 1,2,3 dimensional.
In the absence of the bulk system, the long polymer can access several thermodynamic phases. These phases correspond to distributions over the spatial configurations of the long polymer, $\left\lbrace \vec{r}_{i}\right\rbrace$, where $\vec{r}_{i}$ is the position of the $i^{\text{th}}$ monomer. 
For idealized polymers in regions where boundary conditions can be neglected, these phases can be characterized by scaling relationships between $N_{\text{p}}$ and the radius of gyration $R_{\text{g}}$, defined by $R_{\text{g}}^2 = 1/N_{\text{p}}\sum_i^{N_{\text{p}}}\left(\vec{r}_i - \vec{r}_{\text{cm}}\right)^2$ with $\vec{r}_{cm}$ the center of mass of the chain. We show sample configurations of these phases in Fig.\ref{fig:fig1}B

%(2) Explain the Short extended phase
When $\mu_p$ is negative and $v_{\text{p}}$ is positive or small and negative, the polymer is in the \textit{Short} phase. Here $N_{\text{p}}$ is small, $\approx L$. In our simulations we typically set $\mu_{\text{p}}= 0$, and are not concerned with this \textit{Short} phase.

%(3) Explain the collapsed phase and how you cross over to it from the short extended phase
When $v_{\text{p}}$ is large and attractive, the polymer is in the \textit{Collapsed} phase. Here $N_{\text{p}}$ fluctuates about $N_{\text{max}}$, and typical configurations are space-filling, with a constant density as $N_\text{max} \rightarrow \infty$. 

%(4) explain the long extended phase, transitions between Extended, short, and collpased
When $v_{\text{p}}$ is large and repulsive, but $\mu_p$ is positive, the polymer is in the \textit{Extended} phase. Here the length fluctuates about $N_{\text{max}}$, but the polymer configuration is Extended and the average density of the polymer approaches $0$ as $N_{\text{max}}$ approaches infinity. 

The Extended and Collapsed phase are separated by a line of continuous transitions at $v_{\text{p}}= 0$. We show a typical Extended configuration in Fig.\ref{fig:fig1}B.\\ 

\noindent\textbf{Bulk Molecules Phase Separate and Modify the Polymer Phase Diagram} ---  
%Results 3: Coupling to bulk
%(1) Bulk only phase diagram
Our bulk system has phases and phase-transitions independent of its interactions with the long polymer. The bulk phases are \textit{dilute} and \textit{dense} corresponding to gas and liquid-like states. At certain chemical potentials, or densities, these bulk phases will coexist and the system phase-separates, drawn as the yellow region in Fig.\ref{fig:fig2}E, with typical configurations in each phase shown in Fig.\ref{fig:fig1}C.

%(2) Wetting: depending on interaction, bulk phase will often pull in a collapsed polymer
The polymer configurations in the Wet phase depend on $J_{\text{int}}$ and $J_{\text{bulk}}$, which together determine solvent quality of the dense phase.  
A polymer that prefers to be Extended in an infinite bulk dense phase can be confined by a finite-size droplet separate from the collapse transition.

%(3) prewetting: Existing polymer phases boundaries shift if outside of bulk coexistence
Outside of the bulk coexistence region we find another phase where the polymer is Collapsed and also enriched with a dense-like phase of bulk molecules, shown in Fig.\ref{fig:fig2}B. This is reminiscent of prewetting where a thin bulk domain forms which is adhered to a particular surface phase~\cite{rouches_prewet_21,nakanishi_multicriticality_1982}; we call phases where polymer collapse coincides with polymer-only bulk de-mixing \textit{Generalized Prewet - Collapsed}. Consistent with this view, the polymer phase-diagram effectively shifts and the collapse transition occurs at repulsive $v_{\text{p}}$. Polymer collapse and bulk demixing couple because this minimizes the volume of an otherwise unfavorable bulk phase, while maintaining a high density of bulk-bulk and bulk-polymer interactions. Increasing $\mu_{\text{b}}$, we transition discontinuously between the three phases shown in Fig.\ref{fig:fig2}A-C, which we draw schematically in the phase diagrams sketched in Fig.\ref{fig:fig2}E,F.

%% One paragraph conclusion for the Simulation part? 

\section*{Mean-Field Theory}
Here we develop a mean-field theory that couples a long, collapsible polymer to a bulk fluid with a propensity to phase-separate. 

%(1) Explain terms, explain the ensemble the mft is in, motivate form of F_poly (I Don't know how to derive the form you have here!)
% Fixed length ensemble
\noindent\textbf{Isolated  polymer} --- We write the free energy of an isolated polymer chain, $F^{0}_{\text{poly}}(R_{\text{g}})$, in terms of the order parameter $R_{\text{g}}$, and at fixed length $N_\text{p}$:

\begin{align}
F^{0}_{\text{poly}}(R_{\text{g}}) = &\underbrace{\frac{3}{2}\left(\frac{R_{\text{g}}^2}{N_{\text{p}}a^2} - \log{\frac{R_{\text{g}}^2}{N_{\text{p}}a^2}}\right)}_{\text{Non-Interacting Polymer}} + \underbrace{\frac{v_{\text{p}}}{2}\frac{N_{\text{p}}^2a^3}{R_{\text{g}}^3} +\frac{1}{3!}\frac{N_{\text{p}}^3a^6}{R_{\text{g}}^6}}_{\text{Interactions}}
\end{align}

\begin{figure*}[!htb]
    \centering
    \includegraphics[width=7in]{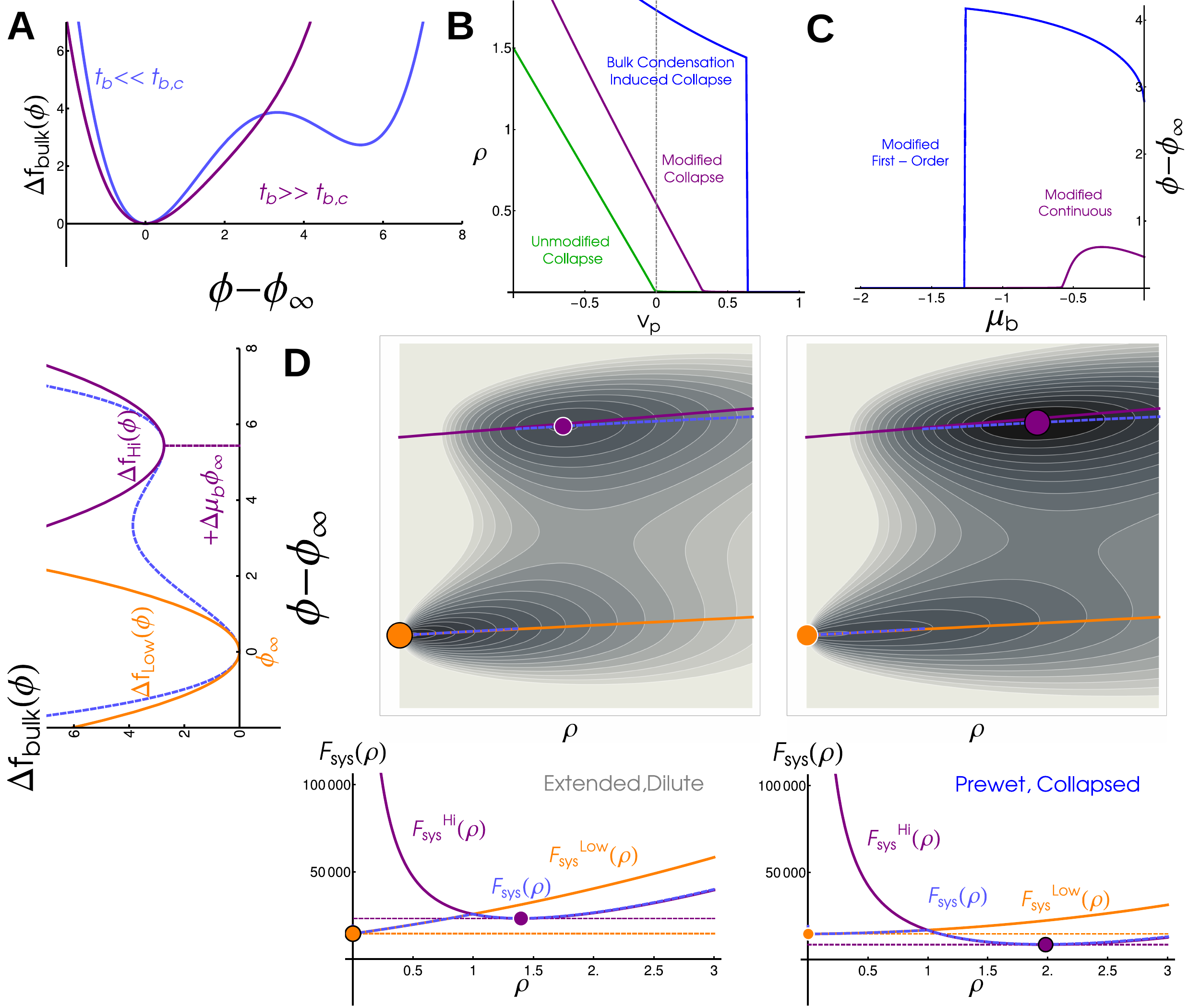}
    \caption{\textbf{Bulk phase-transitions modify the polymer collapse transition in a coupled mean field theory}: (A) Free energy density of the bulk, $\Delta f_{\text{bulk}}$, in super-critical (purple) and sub-critical (blue) regimes. We distinguish these regimes by the presence of one minimum or two well-separated minima. (B) Polymer density $\rho$ as a function of monomer-monomer interaction strength $v_{\text{p}}$. An isolated polymers sees a continuous but sharp collapse transition at $v_{\text{p}}= 0$ (green line), but in coupled systems collapse occurs at repulsive $v_{\text{p}}> 0$ in continuous (purple, bulk has one minima) or discontinuous (blue, bulk has two minima) transitions. (C) Bulk density difference, $\phi - \phi_{\infty}$, as a function of bulk chemical potential $\mu_{\text{b}}$. In coupled systems $\phi-\phi_{\infty}$ increases with $\mu_{\text{b}}$ in continuous (purple) and discontinuous (blue) transitions. In an uncoupled system bulk condensation occurs at $\mu_{\text{b}} = 0$ (not visible on this plot). (D) (left) Quadratic approximation to bulk free energy density in the subcritical regime. Orange corresponds to Dilute phase, purple to Dense phase, and dashed blue is the full free energy. (top) Free energy landscape for a system in the Extended, Dilute phase, $v_\text{p} = 2$. Local minima corresponding to Extended (orange) and Collapsed (purple) phases are marked. The absolute minima of $\phi-\phi_{\infty}$ (dashed, blue) interpolates between the quadratic approximations (solid lines, purple and orange). (bottom) Free energy as a function of polymer density $\rho$. The global minimum of this system is at the low-density, low-$\phi$ minima (orange). (E) Energy landscape (above) and free energy as a function of polymer density (below) for a system in the Collapsed, Prewet phase, $v_\text{p} = 0.1$. The global minimum of this system occurs at the high density, high $\phi$ minima.}
    \label{fig:fig3}
\end{figure*}

The first two terms labeled `Non-Interacting polymer' are entropic contributions to the free-energy from gaussian statistics, and can be derived by replacing the radius of gyration with the end-to-end distance, with $a$ the linear dimension of a single monomer. A free-energy written in terms of $R_{\text{g}}$ has additional entropic costs due to confinement -- but these, along with higher order interactions, are never relevant so we neglect them here (see Supplement). The final two terms capture mean-field interactions between monomers~\cite{de1979scaling}.  The parameter $v_{\text{p}}$ is the difference between monomer-monomer and monomer-solvent interactions. The point $v_{\text{p}}= 0$ separates an Extended polymer phase, where the monomer density $\rho = N a^3/R_{\text{g}}^3$ approaches $0$, from the Collapsed phase where $\rho$ is finite, see the green line Fig.\ref{fig:fig3}B. This transition is continuous but sharp in the large $N_{\text{p}}$ limit ~\cite{de1975collapse,SanchezPhase1979}. \\

%(1) motivate form of F_{\text{b}}ulk, briefly describe its phase diagram
\noindent\textbf{Coupling Polymer to $3D$ Bulk} --- We employ a simple mean field theory for an order parameter $\phi$ which roughly describes the density of bulk phase separating components~\cite{cardy1996scaling,goldenfeld2018lectures}:
%We describe the phase-behavior of a bulk system that de-mixes into two phase with a quartic free energy density~\cite{cardy1996scaling,goldenfeld2018lectures}:
\begin{equation}
\label{eq:bulkf}
f_{\text{bulk}} = \frac{t_{\text{b}}}{2}\phi^2 + \frac{u}{4!}\phi^4 - \mu_{\text{b}}\phi
\end{equation}
%describe the phases of the bulk
Here $t_{\text{b}}$ corresponds to the reduced temperature, a measure of  interaction strength between bulk molecules, $\mu_{\text{b}}$ is the chemical potential, and $u$ captures higher order interactions and is required for stability. %For $t_{\text{b}} < 0$, coexistence occurs at $\mu_{\text{b}} = 0$ with energy minima located at $\phi = \pm \sqrt{6 t_b/u}$. For $t_{\text{b}} \ge 0$ there is one stable phase set by $\mu_{\text{b}}$. These two phases meet in a critical point at $t_{\text{b}} = 0$.  

%(2) motivate form of Fint
In a standard mean field approximation without a polymer, the order parameter is assumed to take the value $\phi_{\infty}$ which globally minimizes the free energy density $f_{\text{bulk}}$.  In the presence of a long polymer, we further assume that a sphere of volume $R_{\text{g}}^3$ has a polymer density $\rho = \frac{N_\text{p}a^3}{R_\text{g}^3}$ and a potentially different value of the order parameter $\phi$.  Here we assume that both order parameters, $R_{\text{g}}$ (or equivalently $\rho$) and $\phi$ together minimize the free energy of the coupled system. 

We describe interactions between the bulk and the long polymer via an energy density $f_{\text{int}}$ and corresponding interaction energy $F_{\text{int}}=f_{\text{int}}R_{\text{g}}^3$:
\begin{equation}
F_{\text{int}}= -h\phi N_{\text{p}}
\end{equation}
%These interactions can be interpreted as the binding energy between the bulk and the long-polymer. 
The parameter $h$ is analogous to $J_\text{int}$, and can be interpreted as a per monomer binding interaction between  bulk and long polymer. To compactly write the coupled free energy of the polymer and its order parameter environment we define $\Delta f_{\text{bulk}} = f_{\text{bulk}}(\phi) - f_{\text{bulk}}(\phi_{\infty})$ and write the free energy of the coupled system as:  
\begin{equation}
\label{eq:Fsys}
F_{\text{sys}}(R_{\text{g}},\phi) = F^0_{\text{poly}}(R_{\text{g}}) + F_{\text{int}}(R_{\text{g}},\phi) + R_{\text{g}}^3\Delta f_{\text{bulk}}(\phi)
\end{equation} 
where we have subtracted off a contribution from the polymer-free bulk.\\

%(3) Discuss integrating out F_int to renormalize parameters of F_poly.
%quadratic potentially -- exactly changes V; goes like susecpetiibility: prewetting, -- polymer co-nonsolvency
\noindent \textbf{A super-critical bulk system modifies the polymer collapse transition} --- To see how the bulk modifies the thermodynamics of the long polymer, we examine the behavior of a weakly interacting bulk by approximating the free energy as quadratic, $\Delta f_{\text{bulk}} = \frac{t_{\text{b}}}{2}\left(\phi - \phi_{\infty}\right)^2$. In this approximation $\phi$ can be minimized analytically, leaving just one additional $R_\text{g}$ dependent contribution to the effective free energy of an isolated polymer, so that $F^{\text{eff}}_{\text{poly}}(R_{\text{g}})=F^0_{\text{poly}}(R_{\text{g}})-\frac{h^2}{t_{\text{b}}}N_{\text{p}}\rho$, and a constant term $-h\phi_{\infty} N_{\text{p}}$.
This additional term has exactly the same form as the monomer-monomer coupling, defining a new $v_{\text{eff}} \equiv v_{\text{p}} - \frac{h^2}{t_{\text{b}}}$. Thus, the effect of the bulk in this regime is precisely to shift the parameter regime of the collapse transition, see purple line in Fig.\ref{fig:fig3}A. The value of the bulk order parameter tracks with the density of the polymer and sees a sharp, continuous increase through the collapse transition, see Fig.\ref{fig:fig3}B, purple line. \\ 

\noindent \textbf{Strongly interacting bulk drives a first-order transition} --- In the limit where the bulk system is instead near to phase coexistence, when $t_{\text{b}} < 0$ the bulk free energy density (equation~\ref{eq:bulkf}) is not well described by a single quadratic minimum.  Still, the order parameters $R_{\text{g}}$ and $\phi$ take values which minimize the combined free energy $F_{\text{sys}}$ (equation~\ref{eq:Fsys}).  %While this minimization cannot be done analytically, minima are easy to find numerically.  
Contour plots of $F_{\text{sys}}$ are shown in Fig.\ref{fig:fig3}D,E (with $\rho$ rather than $R_{\text{g}}$ on the x axis), where there are two distinct minima. When the long polymer interacts more strongly with itself, (lower $v_\text{p}$) there is an abrupt transition as one minimum becomes lower in free energy than the other.  This leads to discontinuities in the polymer density (blue curve in Fig.\ref{fig:fig3}B) and in the bulk order parameter (blue curve in Fig.\ref{fig:fig3}C).

\begin{figure*}[htb]
    \includegraphics[width=6in]{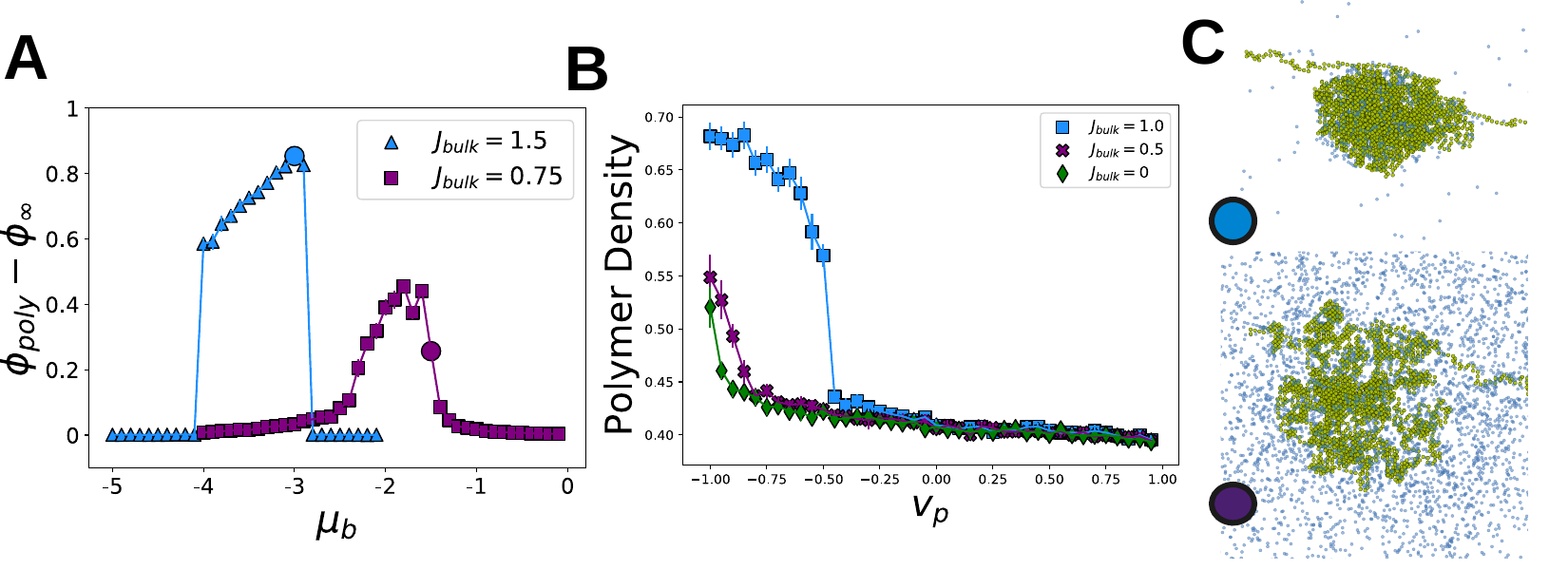}
    \caption{\textbf{Lattice monte-carlo simulations confirm mean-field theory predictions}: --- (A) Density difference between molecules on the polymer and in the bulk as function of $\mu_\text{bulk}$, for three values of $J_\text{bulk}$. For $J_\text{bulk} = 1.0, 1.5$ there is a discontinuous change in the bulk density difference as $\mu_\text{bulk}$ is increased towards coexistence, and an analogous first order transition as the bulk enters its dense phase. For $J_\text{bulk} = 0.5$ there is a steep but continuous increase in the bulk density difference through the transition, and a corresponding transition as the polymer de-condenses; Cf. \ref{fig:fig3}C.  (B) The polymer collapse transition shifts and becomes discontinuous in an interacting solvent. Cf \ref{fig:fig3}B. (C) Simulation configurations corresponding to blue and purple circles in (A). For all simulation parameters, see Supplementary Material} 
    \label{fig:figRev}
\end{figure*}

To gain additional insight into the nature of this abrupt transition, we can also approximate the free energy as quadratic around both dense and dilute phase minima, treating each with the same approximation as above (orange and purple lines in Fig \ref{fig:fig3}D,E left). This predicts that local minima will lie near the corresponding orange and purple lines in Fig ~\ref{fig:fig3}D,E. In this approximation, the system takes the position of either the purple or orange dot, and a first order transition occurs when they have the same value of $F_{\text{sys}}$. Our simulations and theory show that $\rho \rightarrow 0$ as $\mu_{\text{b}} \rightarrow 0$, corresponding to an Extended polymer wet by the bulk dense phase.

Polymer collapse can only be accessed from the extended phase through a phase transition, either continuous or abrupt.  Our simulations are in the abrupt regime, but the presence of a continuous transition above $T_\text{c}$, suggests that a generic phase diagram has the features plotted in figure~\ref{fig:fig2}E; a line of abrupt transitions in the dilute bulk phase but just outside of bulk coexistence, a line of continuous transitions at hight $T_\text{b}$ and near the bulk dense phase, and a tricritical point where they meet. Near the bulk critical point but in the dense phase, a narrow region of prewetting can be accessed through a continuous transition by decreasing bulk density, and thereby increasing effective interactions. This is related to the phenomenon of polymer co-nonsolvency~\cite{mukherji2014polymer}, in which a mixture of a good and bad solvent can induce polymer collapse more effectively than the bad solvent in isolation. 

%stat-mech: as you increase chemical potential the size of the polymer changes. Should go from continuous to critical, but simulations in 1st order regime. 

%%% Confirmation of MFT predictions with our simulations
\noindent\textbf{Monte-Carlo simulations support theoretical predictions} --- Some predictions from our mean-field theory are testable in our Monte-Carlo simulations : (i) Is there a regime of continuous transitions at weak $J_\text{bulk}$, and a regime of discontinuous transitions at large $J_\text{bulk}$? (ii) Do interactions from the bulk solvent shift the collapse transition of the long polymer. We tested these predictions in a minimal Lattice Monte-Carlo Simulation (see Methods). Figure \ref{fig:figRev}A demonstrates a regime where polymer and bulk density increase abruptly as $\mu_\text{b}$ increases for $J_\text{bulk} >> 0$ (blue line), and a regime where the densities continuously change for weaker $J_\text{bulk}$. (purple line). Likewise Figure \ref{fig:figRev}B supports the 'shifting' of the polymer collapse transition with solvent-mediated interactions. This shift can be continuous for weak coupling strengths (purple line) or discontinuous for stronger couplings (blue line). 

The discontinuities we predict will be blunted in finite-size systems. The sharpness of the prewetting-collapse transition is set by the contour length of the long polymer $N_\text{p}$. While biological polymers are larger than most synthetic polymers, a direct mapping between our $N_\text{p}$ and monomer units in biological systems is unclear. Finite size bulk systems can also blunt these discontinuities but only through material depletion, an effect which we do not consider. 

\section*{Multi-component Polymers}
\begin{figure*}[!htb]
    \centering
    \includegraphics[width=7in]{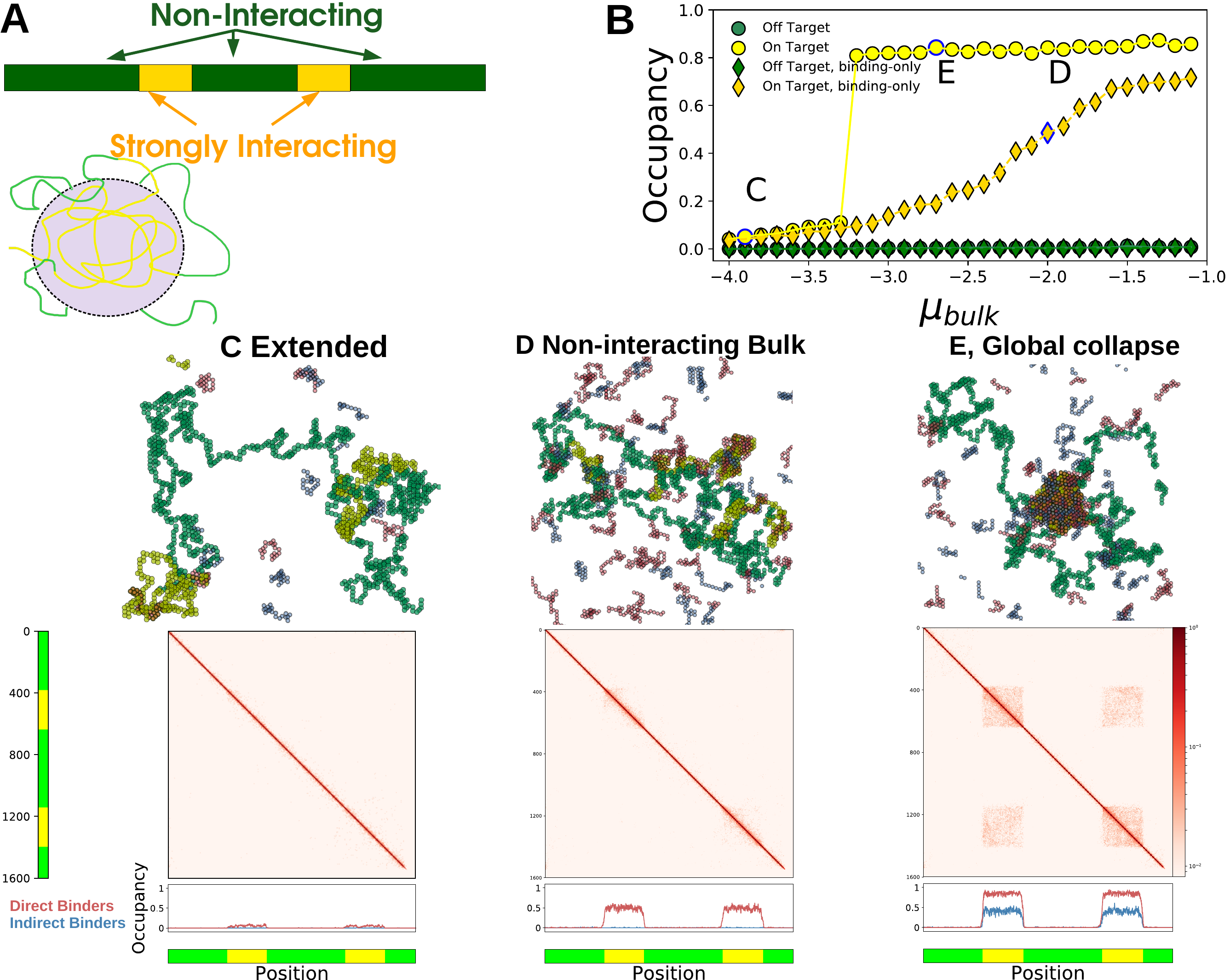}
    \caption{\textbf{Bulk proteins change the configuration of a multi-component polymer} --- (A) (upper) Illustration of a multi-component polymer: the polymer has discrete, contiguous segments of monomers that interact differently with the bulk molecules. Here yellow interacts strongly while green is non-interacting. (lower) Schematic showing the selective collapse of yellow segments into a Prewet phase. (B) Occupancy of yellow and green monomers for the bulk binding molecules, as a function of $\mu_{\text{b}}$. Collapse of the yellow segments coincides a jump in the occupancy of yellow segments, but not green. Letters in the plot correspond to the figures below. (C-E) Simulation configurations, monomer contact probabilities, and occupancy profiles of different multi-component polymer phases. For simulation parameters, see SI (C) Extended phase: the polymer is Extended and bulk occupancy is low and near-uniform. (D) In non-interacting bulk systems, the occupancy of the yellow regions is high (bottom), but the polymer remains in an Extended configuration (middle). (E) Global collapse: (middle) long range, inter-segment contacts are formed via interactions by the bulk phase, red off-diagonal regions in the contact probabilities plot. (lower) Both red and blue bulk molecules are enriched in the yellow segments, but not in the green.}
    \label{fig:fig4}
\end{figure*}
% 1) Simulations and theory can be generalized to multicomponent polymer case. Briefly describe the patterned" polymer. What is different about the monomers, and what the motivation is. 
In many biological scenarios bulk proteins interact strongly with certain regions of the chromosome, and weakly or not at all with others. We modified our simulations by partitioning the long polymer into discrete segments that either interact with the bulk (yellow in Fig \ref{fig:fig4}A) or do not (green in Fig \ref{fig:fig4}A).

% the two basic phases: collapsed and extended
Since the bulk interacts differently with the two monomer types, there are situations where strongly interacting segments collapse but non-interacting segments remain Extended. At low $\mu_{\text{b}}$ the polymer is Extended (Fig ~\ref{fig:fig4}C) while at high $\mu_{\text{b}}$ all interacting segments form a \textit{globally Collapsed} phase (Fig \ref{fig:fig4}E). These phases differ in their occupation of bulk molecules and their polymer-polymer contacts. The Extended phase sees low occupancy of directly binding bulk molecules (red) at the yellow segments (Fig \ref{fig:fig4}C, bottom) while both direct and indirect (blue) bulk molecules are enriched in the globally Collapsed phase (Fig ~\ref{fig:fig4}E, bottom). The occupancy of yellow monomers increases discontinuously with $\mu_{\text{b}}$ while the green monomers remain mostly unoccupied (Fig ~\ref{fig:fig4}B). When bulk molecules do not interact with each other, $J_\text{bulk} = 0$, the occupancy of yellow regions increases smoothly with $\mu_{\text{b}}$ (Fig \ref{fig:fig4}B, diamonds), and there is no accompanying occupancy in indirect binders (Fig \ref{fig:fig4}D, lower).
%% motivate distinction between local and global by differences in polymer-polymer contacts. Discuss how there are transitions between these phases. 
Polymer-polymer contacts in the Extended phase are homogeneous, while the globally Collapsed phase sees long-range contacts between monomers of the same type, even those separated by an intervening sequence (Fig \ref{fig:fig4}E,middle). 
\section*{Discussion}
% a one-pararaph summary
Here we have presented a model where a long polymer near a collapse transition couples to a bulk solvent near a de-mixing transition. We find that this coupling allows both polymer collapse and bulk demixing to occur well outside of their uncoupled phase boundaries. This is reminiscent of prewetting phase-transitions, where a boundary condensed phase is stabilized via interactions with a surface, sometimes undergoing its own transition~\cite{rouches_prewet_21,nakanishi_multicriticality_1982}. Here we discuss the ways biology could use these transitions, and further questions on their physics. \\

\noindent \textbf{Transcriptional activation through a prewetting transition} --- Transcription factors interact directly with enhancers, and many feature `activation domains' that phase-separate with transcriptional machinery such as the co-activator Mediator~\cite{boija2018transcription,SabariCoactivator2018,ChowkMediator2018}, playing roles roughly analogous to the red and blue polymers in our simulations. These Prewet phases could naturally integrate information from transcription factor concentrations into transcriptional regulation by controlling where along the chromosome Prewet domains occur (Fig ~\ref{fig:fig4}B).  Transcription factor concentration could also modify how enhancers, promoters, and distant genomic regions interact in three dimensions, as in Fig~\ref{fig:fig4}E,F. The prewetting phase transition naturally integrates over a long stretch of chromosome, yielding an effective cooperativity in when and where domains appear, as may be relevant for transcription initiation~\cite{chen2018dynamic,levo2022transcriptional,du2024direct,NolisTranscription2009}. 
%\noindent\textbf{Transcriptional repression with Prewet phases }--- Many proteins that interact with heterochromatin and other transcriptionally repressed regions have a propensity to phase-separate in the presence or absence of DNA~\cite{strom2017phase,larson_liquid_2017,quail_force_2021,BantigniesPolycomb2011,kundu2017Polycomb}. Heterochromatin is typically more dense than transcriptionally active chromatin, and is inaccessible to transcriptional machinery~\cite{WangHistone2019}. 
It is possible that heterochromatin is best thought of as a separate generalized Prewet phase, which excludes transcriptional machinery, and is rich in DNA and certain proteins.~\cite{strom2017phase,larson_liquid_2017,quail_force_2021,BantigniesPolycomb2011,kundu2017Polycomb} 

%and read out
%% Interpretation of "regulatory complexes" / CORCs / ChiP
\noindent\textbf{The composition and genomic localization of Prewet domains can encode cellular identity} --- Cellular identity is thought to be encoded by \textit{core regulatory complexes} (CoRCs) composed of a small number of transcription factors which regulate downstream cell-type specific genes~\cite{arendt2016origin,wagner2014homology}.  Evidence for these complexes comes, in part, from Chromatin Immuno-precipitation (ChIP-seq) studies~\cite{Park2016}. Long stretches enriched in cell-type specific groups of a few transcription factors are often interpreted as arising from large, macromolecular complexes. But a generalized Prewet phase localized to certain genomic regions has a similar experimental signature, as Fig.\ref{fig:fig4}E (occupancy) illustrates. Molecules that do not bind directly to the polymer (blue) localize to the same region as molecules that interact directly (red) so long as both of these molecules phase-separate with each other. Many transcription factors that inspired this model for CoRCs phase-separate at high concentrations (OCT4, NANOG ~\cite{boija2018transcription}), and are found as large, localized assemblies in some cell types. Prewet phases which depend on cellular identity may also explain cell type specific differences in the three-dimensional structure of chromatin (Fig.\ref{fig:fig4}E, polymer contacts). \\

\noindent\textbf{Phase diagrams for generalized Prewet phases} --- 
Past work has investigated how binary bulk mixtures can lead to polymer collapse even when either component on its own does not, a phenomena known as polymer-co-nonsolvency~\cite{mukherji2014polymer}. As in that work, far above the critical point in our more complicated bulk mixture, the transition from an Extended to a Collapsed state is continuous, though steep.  Near the two phase coexistence region this transition becomes abrupt, as we see in simulations and mean-field theory ~\ref{fig:fig3}A, compare blue and purple lines.  This implies a tri-critical point which we have not thoroughly explored, near but outside of bulk coexistence, where the line of first-order prewetting transitions meets the line of second-order transitions. We also anticipate that the line of prewetting transitions meets bulk coexistence at another tricritical point, where Dry, Wet, and Prewet phases coexist. At these tricritical points the polymer likely sees non-trivial scaling behavior. We also expect interesting finite-size effects in the wetting regime when the length of the polymer is such that its Collapsed size is of order the size of the bulk domain. And finally, we anticipate coupling many-component bulk systems to a multi-component polymer will yield very rich physics~\cite{graf2022thermodynamic}. \\

%% Comparison to previous work: 1) Phase separatio medieated by (static) polymer (chase), 2) polymer assisted-condensation

% reference Bill, Pierre, Grill? 
\noindent\textbf{Our model differs from past models of phase separation and chromosomes} --- Recent work has investigated the role of polymer `scaffolds' on the phase-separation of $3D$ proteins~\cite{david_phase_2020,sommer2022polymer,morin_sequence-dependent_2022,ancona2022simulating,tortora2023hp1,strom_interplay_2024}, and how $3D$ proteins mediate communication between distal genomic regions~\cite{bialek_action_2019,shin2018liquid, shrinivas_enhancer_2019}. Others have observed that long polymers can widen regimes in which bulk proteins phase separate; here we expand on these observations by studying  the thermodynamics of these transitions
Our results are also related to a body of work investigating how demixing transitions interact with polymer meshes~\cite{ronceray2022liquid,boddeker2022non}, though most of that literature has focused on cytoskeletal meshes, which interact elastically, but do not contribute a transition of their own.  There are regimes where 
%Depending on the sign and strength of the interaction between liquid droplets and mesh components, 
demixing transitions can be enhanced or inhibited, and are generically accompanied by distortions of the polymer mesh. Here we innovate by jointly considering the polymer collapse transition and bulk condensation.  Our work is also related to models of the ParABS system for chromosome segregation in bacteria, which explicitly describe the interactions of a dynamic polymer with interacting proteins, but do not focus on the phase behavior of their bulk system~\cite{broedersz_condensation_2014}. \\ 

%% Why not wetting? %squelching as wetting, maybe only worth a sentence 
\noindent \textbf{Distinguishing prewetting from wetting} --- Although we focus on a region of prewetting phase transitions, our model features other transitions that resemble wetting, where a bulk phase prefers to localize to a surface~\cite{cahn_critical_1977}.  Wet phases are stable even without a long polymer; while polymers can partition into them, the presence of a Wet phase does not imply presence of a specific polymer. As such, Wet phases are less suitable for regulation by controlling DNA's interaction with specific components or its three dimensional structure. Overexpression of transcriptional activators quite generally inhibits transcription through a phenomenon dubbed `squelching'~\cite{gill1988negative}. It is possible that squelching marks a transition into bulk coexistence where macroscopic droplets compete with the DNA for transcriptional machinery. A complementary view is that chromatin-localized condensates serve as heterogeneous nucleation seeds for specific bulk phases~\cite{strom_interplay_2024}. We interpret these 'seeds' as bona-fide prewet phases that may interplay with wet phases in genetic regulation.

%Discussion 3: prewetting is everywhere, -- and we can think of it with the phase-transition in the surface (which others have not done)
Eukaryotic cells have many surfaces with phase-transitions that do not occur in bulk, and biology often needs to localize processes to specific locations. We speculate that biology uses a variety of these \textit{generalized-prewetting} transitions to accomplish specific localization, and to imbue a bulk fluid with properties such as a diverging susceptibility. Work by ourselves and others ~\cite{setru2021hydrodynamic,milovanovic2018liquid,boddeker2022non,morin_sequence-dependent_2022} will continue to uncover the rich physics allowed by surface phase-transitions. \\
\noindent\textbf{Author Contributions} --- M.N.R. and B.B.M. conceived the project. M.N.R carried out the research. M.N.R and B.B.B. analyzed the data and wrote the paper.  

\noindent\textbf{Acknowledgements} --- We thank Asheesh Momi, Yu Fu, Isabella Graf, and Simon Mochrie for feedback on the manuscript, and Pranav Kantroo, Martin Girard, Sarah Veatch, and Simon Mochrie for helpful discussions. This work was supported by NIH R35 GM138341 and NSF 1808551.

\noindent\textbf{Declaration of Interests} --- The authors declare no competing interests. 

\bibliographystyle{unsrt}
\bibliography{refs}

\section*{Materials \& Methods}
\subsection*{Simulation model}
\begin{figure}[!htb]
    \centering
    \includegraphics[width=3.5in]{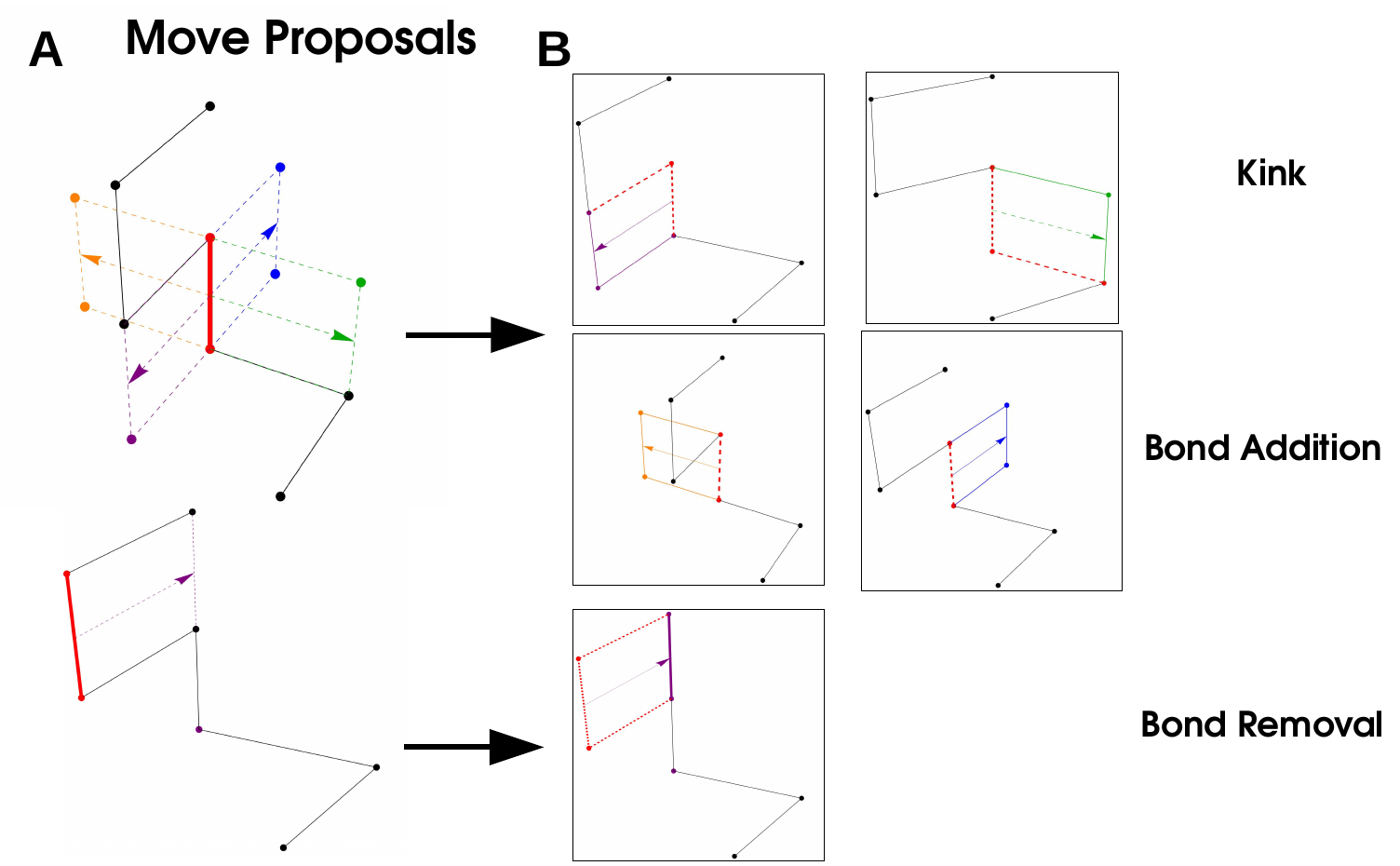}
    \caption{\textbf{Simulation Procedure}: A bond (red) is selected, and translated in a direction orthogonal to its orientation. These moves correspond to a `Kink' (purple and green arrows) where a the polymer configuration changes but the length does not, `Bond Addition' where two monomers are added to the polymer, and  `Bond Removal' where two monomers are removed from the polymer}
    \label{fig:fig5}
\end{figure}

We sample polymer configurations using a Monte-Carlo procedure~\cite{bergPaths1981} composed of three elementary moves that correspond to Addition, Deletion, and Kink of a given bond, where a bond is a link between successive monomers in the polymer, red in Fig ~\ref{fig:fig5} .Any move may be viewed as the translation of a bond in a direction orthogonal to the direction of the bond (arrows in ~\ref{fig:fig5}). The type of move proposals available is completely determined by the vertices of two monomers neighboring the monomers participating in the bond. If translation of the bond gives no intersection with neighboring monomers, the move is \textit{Bond Addition} and corresponds to adding two monomers to the chain. If the translated bond intersects with both monomers in the chain the corresponding move is \textit{Bond Removal} and the polymer loses two monomers. If one of the neighboring monomers is intersected by the translated bond, the move is a \textit{Kink} that does not change the length of the polymer, but alters the configuration. These moves are illustrated in Fig ~\ref{fig:fig5}B. 

We simulate the polymer as follows: we randomly select a monomer, at position $s$, and define its bond as the vector between the monomer and the $s+1$ monomer. We then calculate all bond translations and whether they correspond to addition, removal, or kink moves. We then randomly propose one of the three possible moves, with probabilities:
\begin{align*}
P_{add}(_{\text{p}}) &= \frac{N_{\text{p}}+2}{2*\left(1+N_{\text{p}}\right)}\left(1-P_{kink}\right)\\
P_{rem}(_{\text{p}}) &= \frac{N_{\text{p}}}{2*\left(1+N_{\text{p}}\right)}\left(1-P_{kink}\right)\\
P_{kink}(_{\text{p}}) &= P_{kink}
\end{align*}

These are set to satisfy the detailed balance condition, that $P_{add}(N_p) = P_{rem}(N_p+2)$ and $P_{rem}(N_p) = P_{add}(N_p-2)$; since $P_{kink}$ does not change the length of the polymer it can be held constant. After proposing a move we check for self-avoidance, and that the polymer meets the length constraint $N_p < N_{max}$, if either of these are not satisfied we reject the move. If the move is not rejected, we compute the total energy from $H_{poly}$ and accept with the Metropolis criteria $P_{accept} = max(1,e^{-\beta H_{poly}})$. 

\textbf{Simulation of bulk system}: Simulation of our bulk system closely follows the bulk system used in \cite{rouches_prewet_21}. We simulate a two-component fluid of short polymers whose length $N_b = 20$ monomer units in all simulations.  
We update the position of a single bulk molecule via kink moves, which are analogous to those we propose in the long polymer. Bulk molecules also perform reptation moves where a monomer is removed from one end of the molecule and placed in a position adjacent to the monomer at the other end. These moves conserve length and are proposed randomly with uniform probability. Our bulk is typically held at fixed chemical potential. To this end we separately simulate a reservoir of non-interacting molecules and exchange particles with this reservoir with rate $\lambda_{+} = \frac{N_{r}}{N_{r} + N_{s}+1}$ and  $\lambda_{-} = \frac{N_{s}+1}{N_{r} + N_{s}+1}$ to ensure detailed balance. Particle exchanges and moves are accepted with the Metropolis probability given by the bulk Hamiltonian.

\textbf{Multi-component polymers}: Our simulations of polymers with multiple monomer types use the same move-set as the single-monomer simulations. We first equilibrate a single-monomer polymer for several million MCS, then impose monomer sequence on the polymer. We track the number of monomers per `segment', and moves are allowed to change the monomers-per-segment up to $\pm\Delta$: i.e if we set a segment of 50 monomers of the same type, and $\Delta = 10$, this segment must have $40 -- 60$ monomers. We set $\Delta = 12$ in all simulations displayed in Fig ~\ref{fig:fig4}. Any move that brings the segment outside of this window is rejected. Otherwise the simulations are identical up the Hamiltonian, which now reflects the differences in binding to the bulk
\begin{equation}
\frac{H_{\text{int}}}{k_{\text{B}}T} = -\left(J_{\text{int},1}\sum_{i} s^{1}_{i} s^{\text{p}_1}_{i} + J_{\text{int},2}\sum_{i} s^{1}_{i} s^{\text{p}_2}_{i}\right)
\end{equation}
where, for example, $J_{\text{int},1}$ denotes interactions of the first monomer type with the bulk, and $s^{\text{p},1}_{i}$ denotes the occupancy of the $i^{\text{th}}$ lattice site for monomers of type $1$.

textbf{Minimal Monte-Carlo Simulation}: To validate our mean-field theory, we simulated a simpler bulk system consisting of a single monomeric bulk species and long polymer. The long polymer's Hamiltonian $H_\text{poly}$ and its interactions with the bulk $H_\text{int}$ are unchanged, and we use a lattice-gas Hamiltonian for the bulk:
\begin{align}
\frac{H_{\text{bulk}}}{k_{\text{B}}T} = & -\mu_{\text{b}}\sum_{i} s_{i} - J_{\text{bulk}}\sum_{<i,j>}s_{i}s_{j} 
\end{align}
Here $s_i$ spin variable represents the occupancy of the lattice at site $i^{th}$ for a bulk molecule, and $J_\text{bulk}$ parameterizes a nearest neighbor interaction. Monte Carlo steps for the polymer are unchanged. We propose random flips of bulk spin variables, accepting or rejecting the move with the Metropolis probability.

\textbf{Code Availability}: Source code for the Monte-Carlo Simulations is available on Github: https://github.com/SimludDalhec/PolymerCollapsePrewetting

\end{document}